\documentclass[pra,aps,showpacs,articles,nofootinbib,twocolumn]{revtex4}
\usepackage{epsf,epsfig}
\usepackage[psamsfonts]{amssymb}
\usepackage{amsmath}
\usepackage{color}

\begin{document}

\title{Stationary and uniform entanglement distribution\\ in qubit networks with quasi-local dissipation}

\author{Morteza Rafiee$^{1,2}$\footnote{m.rafiee178@gmail.com},
Cosmo Lupo$^2$\footnote{cosmo.lupo@unicam.it},
Hossein Mokhtari$^1$\footnote{phmh.mokhtari@yazduni.ac.ir} and
Stefano Mancini$^2$\footnote{stefano.mancini@unicam.it}}

\affiliation{$^{1}$Department of Physics, Yazd University,
Pajoohesh St, Safaieh, 89195-741 Yazd, Iran}
\affiliation{$^{2}$School of Science and Technology, University of
Camerino, I-62032 Camerino, Italy}

\begin{abstract}
We consider qubit networks where adjacent qubits besides interacting via $XY$-coupling,
also dissipate into the same environment.
The steady states are computed exactly for all network sizes and topologies,
showing that they are always symmetric under permutation of network sites,
leading to a uniform distribution of the stationary entanglement across the
network.
The maximum entanglement between two arbitrary qubits is shown to depend only
on the total number of qubits in the network, and scales linearly with it.
A possible physical realization by means of an array of doped cavities is
discussed for the case of a linear chain.
\end{abstract}


\maketitle

\section{Introduction}

It has been understood for a long time that entanglement
represents a quintessential and characteristic trait of quantum
mechanics \cite{Schroedinger}. Quantum phenomena are by now very
well known to be key resources for communication and computation
\cite{QCC}, and it has been recently questioned whether they play
a functional role in certain biological processes
\cite{bio_photo,bio_bird}. Due to its fragility under environment
induced decoherence, entanglement is commonly considered to be an
elusive physical phenomenon that can observed only in the most
elementary systems and on the shortest time scales. Nevertheless,
together with a variety of entanglement preserving mechanisms that
have been put forward
\cite{preserv,Julsgaard,ss-ent-p,ss-ent,Polzik,Memarzadeh,BKraus,
Shen}, the idea is now spreading that it can persist on relatively
long time scales, even in a noisy environment, if suitable
conditions are fulfilled. To achieve stationary entanglement in
spin systems it is sufficient to have quasi-local (two-body
interaction) Hamiltonian and local dissipation \cite{ss-ent-p}, or
local Hamiltonian and quasi-local (two-body) dissipation
\cite{ss-ent,Polzik, Memarzadeh}. Till now, these two
possibilities have been studied separately or for systems composed
of a small number of qubits. The main aim of this paper is to
study the effects of both quasi-local interaction and dissipation
in a system composed of an arbitrary number of qubits. Our goal is
to determine general conditions for stationary entanglement and
characterize its distribution among qubits.

We consider a family of models of quantum networks consisting of $n$
qubits with onsite energy and $XY$ interaction between adjacent
qubits. Moreover, a non-Hamiltonian dynamical term is added within the
quantum master equations formalism \cite{ME}. The latter describes
quasi-local dissipation coupling of adjacent qubits, which can
be understood as arising from the coherent damping to the same,
zero-temperature, bosonic environment. We compute the steady
states for any size and network topology.
This allows us to characterize the given model of
quasi-local dissipation as a means for distributing stationary
entanglement over a generic network. We found that the steady
states are largely independent on the dynamical features of the
model, like the strength of the onsite energy or of the $XY$
interaction, and on the network topology. The steady states are
always symmetric under permutation of the network sites, yielding
a uniform distribution of entanglement across the network. In
particular the maximum attainable entanglement between any pair of
qubits, measured by concurrence \cite{Wootters}, equals $2/n$
ebits and is independent on the relative position of the two
qubits. Furthermore, we investigate the steady-state entanglement
as a function of the initial state.
Finally, for the special case in which the network reduces to a
chain we discuss a possible physical realization by means of an
array of doped optical cavities.

The paper proceeds as follows.
In Sec.\ \ref{sec:model} we introduce the model of the qubit network;
in Sec.\ \ref{sec:ss} we compute the steady states of the qubit network;
the distribution of stationary entanglement across the network is discussed
in Sec.\ \ref{sec:ent};
in Sec.\ \ref{sec:polaritons} a possible physical realization is introduced;
Sec.\ \ref{sec:conc} is devoted to conclusions.

\section{The network model}\label{sec:model}

We consider a network of $n$ qubits defined by a connected graph $G$, with vertices $V(G)$
and edges $E(G)$, where a qubit system is sitting at each vertex of the graph and
the edges identify two body interactions between the qubits.
Let
\begin{equation}\label{adjmatrix}
A[G]_{k,l}=\left \{
 \begin{array}{cc}
 0 &\quad{\rm if}\quad k,l \notin E(G)\\
 1 &\quad{\rm if}\quad k,l \in E(G)
  \end{array}
  \right.,
 \end{equation}
be the (symmetric) adjacency matrix of such a graph.

Then, by considering on site energy and $XY$-interaction, the network Hamiltonian
is defined as
\begin{equation}
H = \sum_{k=1}^n \omega_k \sigma_k^\dag \sigma_k
+ \frac{1}{2} \sum_{k\neq l=1}^n [A(G)]_{k,l} J_{k,l} \left( \sigma_k^\dag \sigma_{l} + \sigma_l^\dag \sigma_k \right) \, ,
\end{equation}
where $\omega_k$ and $\sigma_k^{\dag}$, $\sigma_k$ denote
respectively the energy and the raising, lowering operators of the $k$th qubit.
Furthermore $J_{k,l}$ is the coupling strength between qubits $k$ and $l$.

We assume that the dynamics of the network is described by the master equation ($\hbar=1$)
\begin{equation}
\dot\rho = \mathcal{L}(\rho) \, ,
\end{equation}
where
\begin{equation}\label{Liouville}
\mathcal{L}(\rho) = - i [ H , \rho ] + \mathcal{D}(\rho) \, .
\end{equation}
Here, adjacent qubits interact both directly through $H$
and indirectly through a non-Hamiltonian term given in the
Gorini-Kossakowski-Sudarshan-Lindblad form \cite{ME},
\begin{equation}
\mathcal{D}(\rho) = \sum_{k \neq l=1}^n [A(G)]_{k,l} \mathcal{D}_{k,l}(\rho) \, ,
\end{equation}
where
\begin{equation}
\mathcal{D}_{k,l}(\rho) = \frac{\gamma_{k,l}}{2}
\left( 2 L_{k,l} \rho L_{k,l}^\dag - L_{k,l}^\dag L_{k,l} \rho - \rho L_{k,l}^\dag L_{k,l} \right) \, ,
\end{equation}
with $\gamma_{k,l} >0$ and
\begin{equation}
L_{k,l} = \sigma_k + \sigma_{l} \, .
\end{equation}
The non-Hamiltonian term $\mathcal{D}(\rho)$ describes a Markovian damping process
in which pairs of adjacent qubits coherently decay into the same zero-temperature
bosonic bath, with decay rates $\gamma_{k,l}$.

\section{The network steady states}\label{sec:ss}

A pure state $|\psi\rangle \in \mathbb{C}^{2 \otimes n}$ is a
steady state of the network if it satisfies
$\mathcal{L}(|\psi\rangle\langle\psi|)=0$. A characterization of
the pure steady states of an open quantum system undergoing a
Markovian dynamics \cite{ME} has been provided in
\cite{BKraus}. Following \cite{BKraus}, a pure steady state of our
network model is characterized by the conditions:
\begin{enumerate}

\item\label{BK1} $[A(G)]_{k,l} L_{k,l} |\psi\rangle = \lambda_{k,l}
|\psi\rangle$, for all $k$, $l$, with $\lambda_{k,l} \in \mathbb{C}$;

\item\label{BK2} $\left[ i H + \frac{1}{2}\sum_{k \neq l} [A(G)]_{k,l}
\gamma_{k,l} L_{k,l}^\dag L_{k,l} \right] |\psi\rangle = \lambda
|\psi\rangle$, \hspace{3 mm} with $\lambda \in \mathbb{C}$;

\item\label{BK3} $\mathrm{Re}(\lambda) = \frac{1}{2} \sum_{k \neq l}
[A(G)]_{k,l} \gamma_{k,l} |\lambda_{k,l}|^2$, where
$\mathrm{Re}(\lambda)$ denotes the real part of $\lambda$.

\end{enumerate}
To compute the pure steady states of the network we
first notice that the operators $L_{k,l}=\sigma_k + \sigma_l$ are
nilpotent, hence admitting only vanishing eigenvalues.
Thus, condition \ref{BK1} reads
\begin{equation}\label{conds}
[A(G)]_{k,l} \left( \sigma_k + \sigma_l \right) |\psi\rangle = 0 \, .
\end{equation}
A pure state can be expanded in the standard basis,
\begin{equation}
|\psi\rangle = \sum_{a_1, \dots, a_n=0,1} \psi_{a_1, \dots, a_n} |a_1, \dots, a_n \rangle \, ,
\end{equation}
where $\sigma_k^\dag \sigma_k |a_1, \dots, a_n \rangle = a_k |a_1, \dots, a_n \rangle$.
For any pair of adjacent sites, $k$, $l$, Eq.\ (\ref{conds}) implies
\begin{align}
0 = & \sum_{a_k,a_l=0,1} \psi_{a_1, \dots 1_k, \dots, a_l, \dots, a_n} |a_1, \dots, 0_k, \dots, a_l, \dots, a_n \rangle \nonumber \\
& + \psi_{a_1, \dots, a_k, \dots, 1_l, \dots, a_n} |a_1, \dots, a_k, \dots, 0_l, \dots, a_n \rangle \, ,
\end{align}
where the notations $0_k$, $1_k$, $0_l$, $1_l$ are used to indicate
that $a_k=0$, $a_k=1$, $a_l=0$, $a_l=1$, respectively.
This in turn yields
\begin{align}
& \psi_{a_1, \dots, 1_k, \dots, 0_l, \dots, a_n} + \psi_{a_1, \dots, 0_k, \dots, 1_l, \dots, a_n} = 0 \, , \label{cond1} \\
& \psi_{a_1, \dots 1_k, \dots, 1_l, \dots, a_n} = 0 \, .
\end{align}
If there are no isolated points in the network, these conditions
imply that the pure steady states can contain at most one
excitation. They can be written as superposition of the network
vacuum, $|0\rangle \equiv |0_1, \dots, 0_n \rangle$, and the
single excitation states,
$|k\rangle \equiv |0_1, \dots, 1_k, \dots, 0_n \rangle$ for $k=1, \dots, n$.
To simplify the notation we can expand the pure steady states as
\begin{equation}
|\psi\rangle = \alpha |0\rangle + \beta \sum_{k=1}^n \psi_k |k\rangle \, ,
\end{equation}
where the condition (\ref{cond1}) implies
\begin{equation}\label{nnspi}
[A(G)]_{k,l} \left( \psi_k + \psi_l  \right) = 0 \, .
\end{equation}
We can now distinguish two situations according to the network topology:
\begin{itemize}
\item[(i)] If the network does not
contain cycles or it contains only cycles with a even number of
edges, the solution is given by $\psi_k = (-1)^{n_k} \psi_1$,
where $n_k$ is the number of edges connecting the $k$th site with
the first one;
\item[(ii)] Otherwise, if the network contains cycles with
odd number of edges, the only solution is obtained by putting
$\psi_k=0$ for all $k$.
\end{itemize}
In conclusion, we get that the pure steady states have the form
\begin{equation}
|\psi\rangle = \alpha |0\rangle + \beta |\aleph\rangle \, ,
\end{equation}
where
\begin{equation}\label{aleph}
|\aleph\rangle = \frac{1}{\sqrt{n}} \sum_{k=1}^n (-1)^{n_k} |k\rangle \, .
\end{equation}
The coefficients $\alpha,\beta\in\mathbb{C}$ are arbitrary if the
network topology fulfills (i).
On the other hand, we have to put $\beta=0$ if (ii) holds.
Being interested in the distribution of stationary entanglement, in
the following we assume that (i) is verified.

Then, the condition \ref{BK2} reads
\begin{equation}
i H \left( \alpha |0\rangle + \beta |\aleph\rangle \right) = \lambda \left( \alpha |0\rangle + \beta |\aleph\rangle \right) \, ,
\end{equation}
which may have two independent solutions:
\begin{itemize}

\item
The first solution is obtained for $\alpha=0$ under the conditions
\begin{equation}\label{omegaJ}
i\lambda = \omega_k - \sum_{l} [A(G)]_{k,l} J_{k,l} \, ,
\end{equation}
for all $k=1, \dots, n$.

\item
The second solution is obtained for $\beta=0$, with $\lambda=0$;

\end{itemize}
In the degenerate case, $\omega_k-\sum_{l} [A(G)]_{k,l} J_{k,l}=0$, there exists
a two-dimensional steady subspace $\mathcal{H}_s = \mathrm{span}\{ |0\rangle, |\aleph\rangle \}$.
Otherwise if $\omega_k-\sum_{l} [A(G)]_{k,l} J_{k,l} = const. \neq 0$, the only pure steady
states are $|0\rangle$ and $|\aleph\rangle$.
Finally, we notice that the condition \ref{BK3} is satisfied in
both cases, since $\mathrm{Re}(\lambda)=0$.

Furthermore, it is worth noticing that it could be possible that
other mixed steady states exist, which are not in the convex
hull of pure steady states.

The steady states of our models do fulfill $[A(G)]_{k,l} L_{k,l}|\psi\rangle=0$.
The steady states satisfying such a property are called {\it dark states}.
A uniqueness theorem for the dark states has been provided in \cite{BKraus},
for a system admitting a subspace of dark states.
We can hence apply this theorem in the degenerate case, in which the
subspace $\mathcal{H}_s = \mathrm{span}\{|0\rangle,|\aleph\rangle\}$ is
a subspace of dark states.
According to this result, if there exists no subspace $S$ with
$S \perp \mathcal{H}_s$ such that $[A(G)]_{k,l} L_{k,l} S \subseteq S$, then
the only mixed steady-states are in the convex hull of $\mathcal{H}_s$.
It is easy to show that a subspace with such a property does not exist
for our models. To show that it is sufficient to notice that since the
operators $L_{k,l}$ are nilpotent, a subspace $S$ which is invariant
under the action of all the $L_{k,l}$'s must necessarily include
$|0\rangle$ or $|\aleph\rangle$. Thus, $S$ cannot be orthogonal
to $\mathcal{H}_s$.

\section{Steady-state entanglement distribution}\label{sec:ent}

In this section we study the steady-state entanglement between two
arbitrary network qubits. Due to the symmetric form of
(\ref{aleph}), the extension of the analysis of multiqubit
entanglement is straightforward.

Since we are interested in the distribution of steady-state entanglement
across the network, in the following we will restrict to the case in which
both condition (ii) and Eq.\ (\ref{omegaJ}) are fulfilled.
Under this hypothesis, the reduced steady state of two arbitrary qubits at
site $k$ and $j$ of the network is necessarily of the form
\begin{equation}
\rho_s^{k,j} = \left( 1 - \frac{2p}{n} \right) |00\rangle\langle 00| + \frac{2p}{n} |\Psi^{k,j}\rangle\langle\Psi^{k,j}| \, ,
\end{equation}
where $|\Psi^{k,j}\rangle = \left[ |10\rangle + (-1)^{n_j-n_k}
|01\rangle \right]/\sqrt{2}$ is a maximally entangled state.
It is worth noticing that the reduced steady-state contains only one
free parameter, $p$.
Such a parameter is determined by the initial state of the network.
To fix the ideas, we consider the concurrence \cite{Wootters} as an
entanglement measure for the reduced state of the two qubits.
The stationary concurrence of the two-qubit reduced state is as well
a function of $p$ and the total number of qubits, namely
\begin{equation}
C_s = \frac{2p}{n} \, .
\end{equation}

In order to evaluate $p$ for a given initial state of the network, let us notice that
\begin{equation}
\langle \aleph | \dot\rho(t) | \aleph \rangle = \frac{1}{2} \sum_{k \neq l=1}^n [A(G)]_{k,l} \gamma_{k,l} \langle \aleph | L_{k,l} \rho(t) L_{k,l}^\dag | \aleph \rangle \, ,
\end{equation}
where we have used the fact that $[A(G)]_{k,l} L_{k,l}|\aleph\rangle=0$,
and that the vectors $L_{k,l}^\dag | \aleph \rangle$ are superpositions of states
containing two or more excitations. Moreover, we remark that the
total number of excitations in the network cannot increase under
the evolution dictated by the master equation (\ref{Liouville}).
Therefore, we conclude that if the initial state $\rho(0)$ contains up to one excitation, it
follows that $\langle \aleph | L_{k,l} \rho(t) L_{k,l}^\dag | \aleph
\rangle = 0$ for any $t \geq 0$, which in turn yields that the quantity
$\langle \aleph | \rho(t) | \aleph \rangle$ is a constant of motion.
Eventually we get $p=\langle \aleph | \rho(0) | \aleph \rangle$, which
allows us to compute the steady-state parameter $p$ for any initial state of
the network, provided it contains up to one excitation.
Let us further explore this setting by assuming that the
network is initialized in a state containing a single excitation
over $m$ qubits, that is, $|\psi(0)\rangle = \sum_{j=1}^m \alpha_j |k_j\rangle$.
The maximum stationary concurrence of the two-qubit reduced state
is hence obtained by maximizing $p=|\langle \aleph | \psi_0\rangle|^2$.
It follows that the optimal choice for the initial state is
\begin{equation}\label{optm}
|\psi(0)\rangle = \frac{1}{\sqrt{m}} \sum_{j=1}^m  (-1)^{n_{k_j}} |k_j\rangle \, ,
\end{equation}
yielding $p=m/n$.

We define $C_s(1,m)$ as the maximum stationary concurrence that can
be achieved by preparing the network into an initial state containing up to $1$
excitation over $m$ qubits. We then have obtained that
\begin{equation}\label{concur}
C_s(1,m) = \frac{2m}{n^2} \, .
\end{equation}
To go beyond the single-excitation setting, we have analyzed numerically the
achievable stationary concurrence for initial states containing more than
one excitation. By defining $C_s(N,m)$ as the maximum stationary concurrence
for an initial state containing up to $N$ excitations over $m$ qubits, our
numerical investigations suggest to conjecture that
\begin{equation}
C_s(N,m) \leq C_s(1,m) \, ,
\end{equation}
where the optimal network initial state is as in Eq.\ (\ref{optm}), that is,
a single-excitation initial state is sufficient to achieve the overall
maximum concurrence for a given $m$.

\section{Physical realization}\label{sec:polaritons}

We sketch here a possible physical realization by an array of $n$ doped cavities coupled via
optical fibers (see, e.g., \cite{Lepert,Angelakis}).
Actually, we restrict our attention to the case in
which the network is a linear chain with open boundary conditions.
The case of periodic boundary conditions can be analyzed in a
similar way. In the case of a linear chain, the model generalizes
that introduced in \cite{Memarzadeh} where the Hamiltonian term is
dropped.

Each cavity is doped with a two-level atom and is coupled by optical
fibers to the next-nearest cavities. We denote as $c_k$,
$c_k^\dag$ the ladder operators of the $k$th cavity, coupled to
the levels $|g\rangle_k$, $|e\rangle_k$ of the $k$th atom.
Neighboring cavities are in turn coupled through a single fiber
mode, having ladder operators $a_k$, $a_k^\dag$. Furthermore, we
assume that the $k$th fiber mode interacts with its bosonic
environment, described by a collection of operators $\{ b_{k,j},
b_{k,j}^\dag \}$. The Hamiltonian of the system in the rotating
wave approximation is given by
\begin{equation}\label{hamltonian}
H = H^{free} + H^{int} \, ,
\end{equation}
where
\begin{align}
H^{free} = & \sum_{k=1}^n \omega_k^c c_k^\dagger c_k + \sum_{k=1}^n \omega_k^a |e\rangle_k\langle e| + \sum_{k=1}^{n-1} \omega_k^f a_k^\dagger a_k \nonumber \\
         & + \sum_{k=1}^{n-1} \sum_j \omega_{k,j}^e b_{k,j}^\dagger b_{k,j} \, ,
\end{align}
and
\begin{align}
H^{int} = & \sum_{k=1}^n f_k \left( c_k^\dag |g\rangle_k\langle e| + H.c. \right) \nonumber \\
        & + \sum_{k=1}^{n-1} J_k \left[ a_k^\dagger \left( c_k + c_{k+1} \right) + H.c. \right] \\
        & + \sum_{k=1}^{n-1} \sum_j \eta_{k,j} \left( a_k^\dagger b_{k,j} + H.c. \right) \, . \nonumber
\end{align}
The first and second term in $H^{free}$ are the free Hamiltonian of the
cavity field and the two level atom inside each cavity, the third
and forth term describe the free Hamiltonian of the fibers modes and of
the environment of each fibers with mode frequencies $\omega_k^c$,
$\omega_k^a$, $\omega_k^f$ and $\omega_k^e$ respectively.
Also, the first term in the $H^{int}$ describes the interaction between
the cavity mode and the atom inside the cavity with the coupling strength
$f_k$, the second term is the interaction between the cavity and
the fibers modes with the coupling strength $J_k$ and the last term is the
interaction between the fibers and their bosonic baths with the
coupling strength $\eta_{k,j}$. To write the above Hamiltonian we assumed
the cavities are in the strong coupling regime, i.e., $f \gg
\kappa^a, \kappa^c$, where $\kappa^a$ and $\kappa^c$ are the
atomic and cavity decay rates respectively. So, we assume that
both the decay rates are negligible compared with the coupling
between the fibers and their environments.

The first two terms of $H^{free}$ and the first term of $H^{int}$
can be jointly diagonalized in the basis of polaritons
\cite{Angelakis}. On the resonance between atom and cavity, i.e.,
$\omega_k^c=\omega_k^a\equiv\omega_k$, the polaritonic states
$|n\pm\rangle_k=(|g,n\rangle_k\pm|e,n-1\rangle_k)/\sqrt{2}$, with
energies $E_k^\pm=n\omega_k\pm f_k\sqrt{n}$, are ``created'' by
the operators $P_k^{(n \pm)\dagger}=|n\pm\rangle_k \langle g,0|$.
Due to photon blockade and in the Mott phase, double or higher
occupancy of the polaritonic states is prohibited, hence the only
states to be considered are $|1,\pm\rangle_k$, with energies
$\omega_k \pm f_k$ \cite{Birnbaum, Imamoglu}. Moreover, in
rotating-wave approximation and interaction picture the
inter-converting terms between different polaritons
$P_k^{(1-)\dag}P_{k+1}^{(1+)}$ and $P_k^{(1+)\dag}P_{k+1}^{(1-)}$)
in the interaction Hamiltonian are fast rotating and they average
to zero. So, if initially polaritons are created solely by
$P_k^{(1-)\dag}$, which is possible by applying a global external
laser to the atom-cavity system [16], the polaritonic state
$|1,+\rangle$ will never appear. Then Hamiltonian
(\ref{hamltonian}), taking into account that each polariton can be
treated as a two level system with ladder operator
$\sigma_k^\dagger \equiv |1,-\rangle_k \langle g,0|$, can be
rewritten as
\begin{align}
H = & \sum_{k=1}^n (\omega_k-f_k) \sigma_k^\dagger \sigma_k +
\sum_{k=1}^{n-1} \omega_k^f a_k^\dagger a_k
+ \sum_{k=1}^{n-1} \sum_j \omega_{k,j}^e b_{k,j}^\dagger b_{k,j} \nonumber \\
  & + \sum_{k=1}^{n-1} J_k \left[ a_k^\dagger \left( \sigma_k + \sigma_{k+1} \right) + H.c. \right] \\
  & + \sum_{k=1}^{n-1} \sum_j \eta_{k,j} \left( a_k^\dagger b_{k,j} + H.c. \right) \, . \nonumber
\end{align}

By adiabatic elimination of the fibers mode operators we obtain
the effective Hamiltonian
\begin{align}
H^{eff} = & \sum_{k=1}^n \omega'_k \sigma_k^\dagger \sigma_k + \sum_{k=1}^{n-1} \sum_j \omega_{k,j}^{\prime e} b_{k,j}^\dagger b_{k,j} \nonumber \\
        & + \sum_{k=1}^{n-1} J'_k \left( \sigma_k^\dagger \sigma_{k+1} + \sigma_{k+1}^\dagger \sigma_k \right) \\
        & + \sum_{k=1}^{n-1} \sum_j \eta'_{k,j} \left[ b_{k,j}^\dagger ( \sigma_k + \sigma_{k+1} ) + H.c. \right] \, , \nonumber
\end{align}
with
\begin{align}
\omega'_k           & = \omega_k - f_k - \frac{2J_k^2}{\omega_k^f} + \frac{J_k^2}{\omega_k^f} \delta_{k,1} + \frac{J_k^2}{\omega_k^f}\delta_{k,n} \, , \\
J'_k                & = - \frac{J_k^2}{\omega_k^f} \, , \\
\omega_k^{\prime e} & = \omega_k^e - \frac{2J_k^2}{\omega_k^f} \, , \\
\eta'_{k,j}         & = - \frac{J_k \eta_{k,j} }{\omega_k^f} \, .
\end{align}
This Hamiltonian describes a qubit chain with $XY$ interaction
where nearest-neighbor qubits dissipate into the same bosonic
bath. By tracing out the bosonic baths, which are assumed at
zero-temperature, and in the Born and Markov approximations, one
finally obtains the master equation (\ref{Liouville})
describing the polariton system.

\section{conclusion}\label{sec:conc}

We have presented a characterization of the steady-states of qubit
networks where adjacent qubits are coupled both directly
via a $XY$ interaction, and indirectly via the coherent
dissipation into the same bosonic bath at zero temperature.
We have determined conditions allowing the distribution of
steady-state entanglement. Rather interesting, the features of the
steady-state entanglement are largely independent on the network
topology and on the dynamical details (e.g., coupling constants and decay rates).
The maximal amount of steady-state entanglement that can be achieved between two
arbitrary qubits only depends on the size of the network and
decreases linearly with it.
The steady-state entanglement is also a function of the network
initial state. Furthermore, our analytical results, supported by
numerical evidences leads us to conjecture that the optimal initial
state of the network is a symmetric, Dicke-like, state containing
a single excitation.

An array of doped optical cavities coupled by
optical fibers is also discussed as a physical implementation,
at least for the case of a network reducing to a linear chain. Another system could be
that of planar arrays of trapped electrons used for quantum
information processing \cite{Penning}.

Finally, the performed study lends itself to
consider extension from 2-body dissipation to $n_d$-body dissipation in $n$
qubits network and to analyze the scaling properties of entanglement
vs $n_d/n$. This is left for future explorations.

\acknowledgments

%We acknowledge the financial support of the European
%Commission, under the FET-Open Grant Agreement HIP,
%Grant No. FP7-ICT-221889.

The authors thanks L. Memarzadeh for useful discussion and
comments. MR would like to thank the Ministry of Science, Research
and Technology of Iran for the financial support and also the
University of Camerino for the kind hospitality. The research
leading to these results has received financial support from the
European Commission, under the FET-Open Grant Agreement HIP, Grant
No. FP7-ICT-221889.

\end{document}